# Spectral imaging of thermal damage induced during microwave ablation in the liver*

Neil T. Clancy, Kurinchi Gurusamy, Geoffrey Jones, Brian Davidson, Matthew J. Clarkson, David J. Hawkes and Danail Stoyanov

*Abstract*— **Induction of thermal damage to tissue through delivery of microwave energy is frequently applied in surgery to destroy diseased tissue such as cancer cells. Minimization of unwanted harm to healthy tissue is still achieved subjectively, and the surgeon has few tools at their disposal to monitor the spread of the induced damage. This work describes the use of optical methods to monitor the time course of changes to the tissue during delivery of microwave energy in the porcine liver. Multispectral imaging and diffuse reflectance spectroscopy are used to monitor temporal changes in optical properties in parallel with thermal imaging. The results demonstrate the ability to monitor the spatial extent of thermal damage on a whole organ, including possible secondary effects due to vascular damage. Future applications of this type of imaging may see the multispectral data used as a feedback mechanism to avoid collateral damage to critical healthy structures and to potentially verify sufficient application of energy to the diseased tissue.**

## I. Introduction

Microwave ablation (MWA) is a minimally invasive technique that enables destruction of tissue, such as tumours, *in situ* through thermal damage. This is achieved by absorption of microwave energy by water, delivered by one or more 'needle' electrodes, over a period of several minutes. MWA is part of a group of similar technologies, such as radiofrequency ablation, cryoablation, high intensity focused ultrasound, and recently irreversible electroporation ablation, that all aim to destroy tumour cells, while minimizing damage to surrounding healthy tissue. However, there remains a degree of subjectivity in the application of the technique, and monitoring the zone of thermal damage is not practically possible with commonly-used medical imaging techniques (e.g., ultrasound). Recent results suggest that the ablation needle may even have systemic effects [1] in addition to the desired local damage. Thermal imaging provides some estimates of the energy deposition in the tissue but it is not clear how this translates to physical effects on organ health. Further data on how tissue properties change with application of MWA would increase understanding of how it could be applied clinically, decreasing the risk to healthy structures.

Optical spectral imaging techniques, both multispectral (MSI) and hyperspectral (HSI), are capable of monitoring functional and structural changes in tissues intraoperatively in a non-invasive manner, measuring oxygen content, as well as heterogeneities in physical structure. Deep-tissue imaging techniques such as photoacoustic tomography (PAT) and diffuse correlation spectroscopy (DCS) can detect vascular dynamics >1 cm under the surface and have seen recent commercial development [2]. Both have practical limits: PAT requires direct tissue contact and a complex laser source, and DCS is subject to motion errors. MSI has the advantage of providing high spatial resolution over a wide field-of-view, whilst using standard clinical light sources. Depending on the type of MSI camera and the number of wavelengths required, framerates of <1 Hz to 30 Hz can be achieved. Preliminary work has also reported the use of normal RGB sensors to compute surrogates to MSI results [3, 4]. Previous work has applied the technique in studies of tissue oxygen saturation ($StO_2$) in the bowel and uterus [5, 6]. We have also reported the use of spectral data to map changes in tissue structure after application of radiofrequency tissue fusion [7, 8].

In this paper we describe the results of a surgical experiment where the porcine liver was imaged using MSI and thermal cameras during application of MWA. The time course of the tissue temperature rise and corresponding $StO_2$ change is measured, along with an analysis of the differences of the spatial extent of the observed effects.

## II. Materials and Methods

### A. Microwave Ablation in Porcine Liver

A laparotomy was performed during an open porcine procedure (Yorkshire pig; ~45 kg). The liver was identified and exposed, and a microwave (2.45 GHz) ablation needle (Solero MTA; AngioDynamics, Inc., USA) introduced under direct vision. The ceramic tip, where energy is delivered, was approximately 1 cm below the surface of the tissue. Baseline readings of the tissue were recorded for approximately 30 s, before initiation of ablation, which involved delivering 133 W of energy over a period of four minutes. Data acquisition was terminated approximately 30 s after ablation ceased.

### *B. Imaging and Sensing*

During the ablation procedure optical measurements were made continuously with both a thermal (A35sc; FLIR Systems, Inc., USA) and a multispectral camera (SpectroCam;

*This work is funded by a Wellcome Trust Pathfinder award (201080/Z/16/Z) and by the EPSRC (EP/N013220/1, EP/N022750/1, EP/N027078/1, NS/A000027/1, EP/P012841/1).

N. T. Clancy, G. Jones and D. Stoyanov are with the Wellcome/EPSRC Centre for Interventional & Surgical Sciences (WEISS), Centre for Medical Image Computing (CMIC), and Department of Computer Science, all at University College London, WC1E 6BT, UK. (corresponding author: n.clancy@ucl.ac.uk; geoffrey.jones.12@ucl.ac.uk; danail.stoyanov@ucl.ac.uk).

M. J. Clarkson and D. J. Hawkes are with the Wellcome/EPSRC Centre for Interventional & Surgical Sciences (WEISS), Centre for Medical Image Computing (CMIC), and Department of Medical Physics and Biomedical Engineering, University College London, WC1E 6BT, UK (m.clarkson@ucl.ac.uk; d.hawkes@ucl.ac.uk)

K. Gurusamy and B. Davidson are with the Division of Surgery and Interventional Science, UCL Medical School, Royal Free Hospital, University College London, NW3 2QG, UK. (k.gurusamy@ucl.ac.uk; b.davidson@ucl.ac.uk).

Pixelteq, USA). This camera has eight filter slots, which can be customized to suit the application. The configuration used in this experiment is shown in Table1. A fibre optic diffuse reflectance spectroscopy (DRS) probe (RP22; Thorlabs Ltd., UK) coupled to a tungsten halogen light source (SLS201L/M; Thorlabs Ltd., UK) and CCD spectrometer (CCS200; Thorlabs Ltd., UK) was used to measure high resolution reflectance spectra at several locations on the organ during the baseline and post-ablation phases. Spectral accuracy of the camera was confirmed by making parallel measurements of a standardized colour checker card (Spyder ColorCHECKR24; Datacolor, Inc., USA) with both SpectroCam and the reflectance probe. The error in SpectroCam's measurements is quoted as the mean RMS difference between the spectra.

TABLE I. . FILTER CONFIGURATION FOR SPECTROCAM MSI. EACH FILTER HAS A GAUSSIAN PROFILE, AND PASSBAND SPECIFIED BY FULL-WIDTH AT HALF-MAXIMUM (FWHM).

| **Wavelength (nm)** | 470 | 480 | 520 | 540 | 560 | 578 | 615 | 645 |
|---|---|---|---|---|---|---|---|---|
| **FWHM (nm)** | 20 | 25 | 20 | 18 | 20 | 10 | 17 | 17 |

The experimental set-up involved placement of the cameras above the surgical incision so that the organ filled the field of view. The MSI camera was mounted on an adjustable arm, fixed to the operating table, while the thermal camera was mounted on a separate stand with a top-down view. This allowed visualization of the organ with minimal disruption to movement of surgeons and other theatre staff.

## III. DATA PROCESSING

MSI and DRS reflectance spectra were calculated by dividing measured intensities by a reference spectrum of a known standard (Spectralon SRS-99-010; Labsphere, Inc., USA), allowing further spectral comparisons. Further processing to calculate absorbance and oxygen saturation ($StO_2$) was carried out according to a method described previously [5]. This involved taking the negative logarithm of reflectance to calculate absorbance and applying a linear regression model of light attenuation in tissue. This approach assumes that 1) haemoglobin is the only absorber in the field of view, 2) scattering is constant across the spectrum, and 3) the penetration depth of light into the tissue is the same for all wavelengths. Pure component spectra (oxygenated and deoxygenated haemoglobin) were obtained from a widely-used dataset [9] and convolved with the transmission spectra of the MSI camera's filters (Table 1). For the imaging results, pixel locations that were a poor match to the regression model ($r^2 < 0.9$) were excluded from quantitative analysis.

The spectral resolution of the spectrometer used here was approximately 2 nm, whilst the resolution of the MSI camera varied between filters and was significantly larger (17-25 nm). Therefore, in order to make meaningful comparisons between the two, the output of the point probe was convolved with the transmission spectra of each SpectroCam filter and plotted against their corresponding centre wavelengths.

## IV. RESULTS

Fig. 1 shows the results of the spectral validation, and indicates good agreement between the modalities across the visible range. Quantitatively, there was a mean RMS error of 2%.the bottom row corresponds to the white and grey panels of the colour card, which all appear flat and at the same level, due to the normalization step.

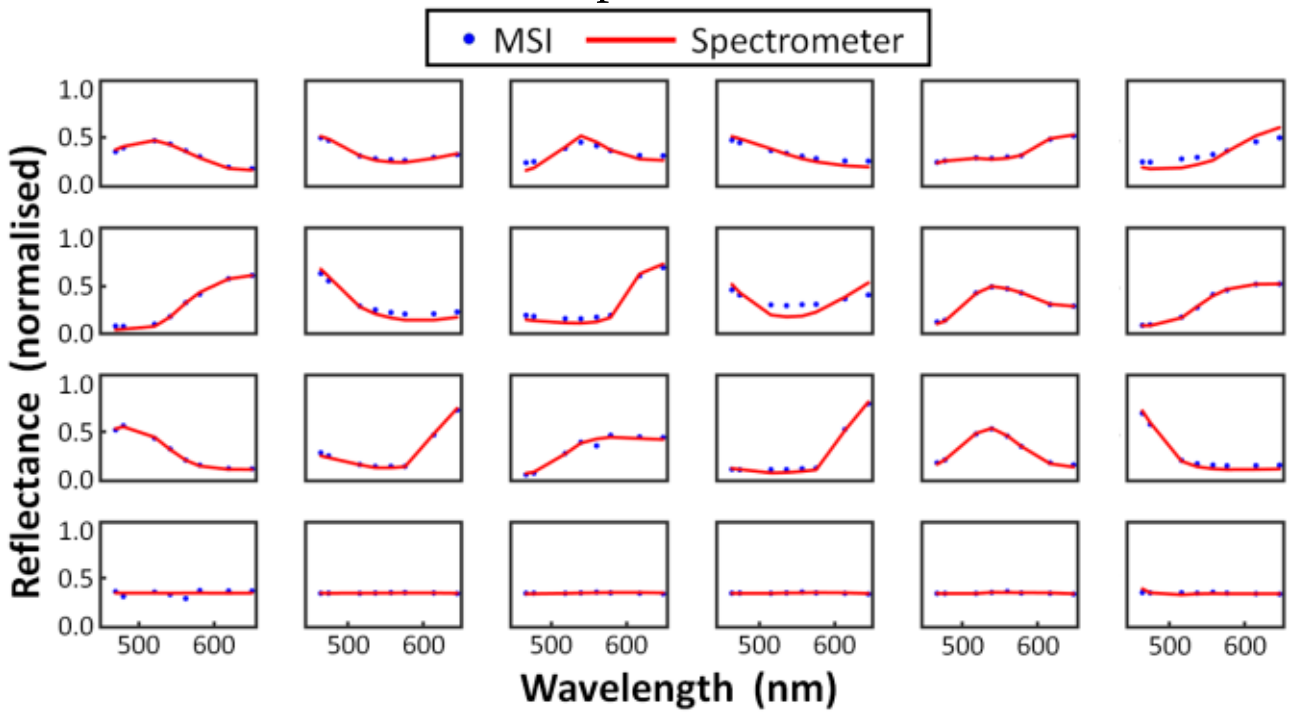

Figure 1. Validation of SpectroCam MSI reflectance measurements against high resolution reflectance spectrometer results. Each subplot shows normalised reflectance plotted against wavelength for the corresponding subpanel in the SpyderChecker card.

Fig. 2 shows a time series of the ablation consisting of RGB, $StO_2$ and thermal images as MWA is performed. The RGB images on their own reveal macroscopic changes in the organ during energy delivery, with the most intense changes visible close to the needle where the tissue appears to turn pink, then eventually black and a region of coagulation necrosis becomes apparent. Outside this central necrotic area is a zone of brighter pink, while a further uneven discoloured region is seen to spread from the lower left corner.

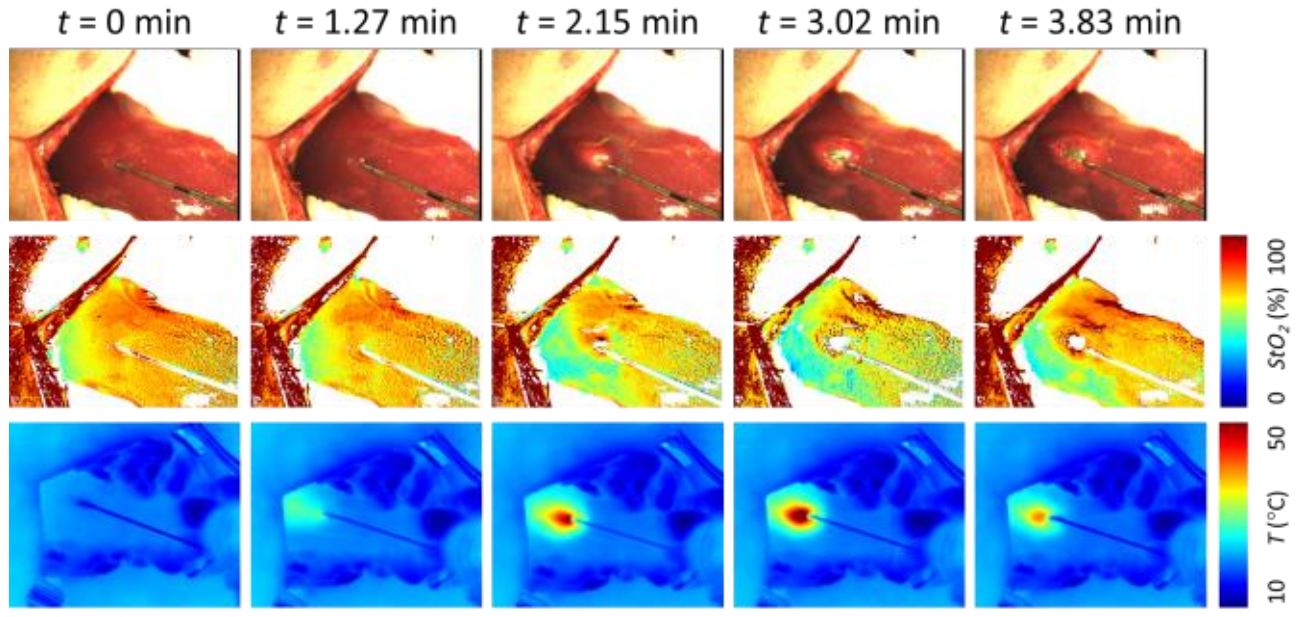

Figure 2. Image series showing time course of MWA in the liver. Top row: RGB reconstructed from three SpectroCam filtlers. Middle row: oxygen saturation images. White areas correspond to pixels with a poor match between modelled and measured reflectance. Bottom row: thermal images showing temperature rise in tissue close to the tip of the needle.

The thermal camera images show the zone of thermal effect spread evenly, symmetric about the needle itself. The central zone reaches a peak temperature of greater than 60°C from a baseline value of 19°C.

Fig. 3 shows a series of regions-of-interest (ROIs) in the $StO_2$ images, along with their corresponding mean values, plotted along a line to indicate how the profile of oxygen saturation varies during the ablation. The $StO_2$ colour maps show that the baseline tissue oxygen saturation in the liver is generally high, varying between 80-88% across the organ. As MWA is initiated this begins to change, with the most noticeable effects again occurring close to the needle. Increased specular highlights in the central region prevent use of the regression algorithm, and subsequent calculation of $StO_2$ as seen in Fig. 3, but the surrounding region shows increased oxygenation (5-10%). This is an effect that becomes less prominent with increasing distance from the treatment zone. However, the ischaemic zone visible in the bottom of the images does not appear to follow this trend, and its asymmetric shape does not follow that of the zone of temperature increase.

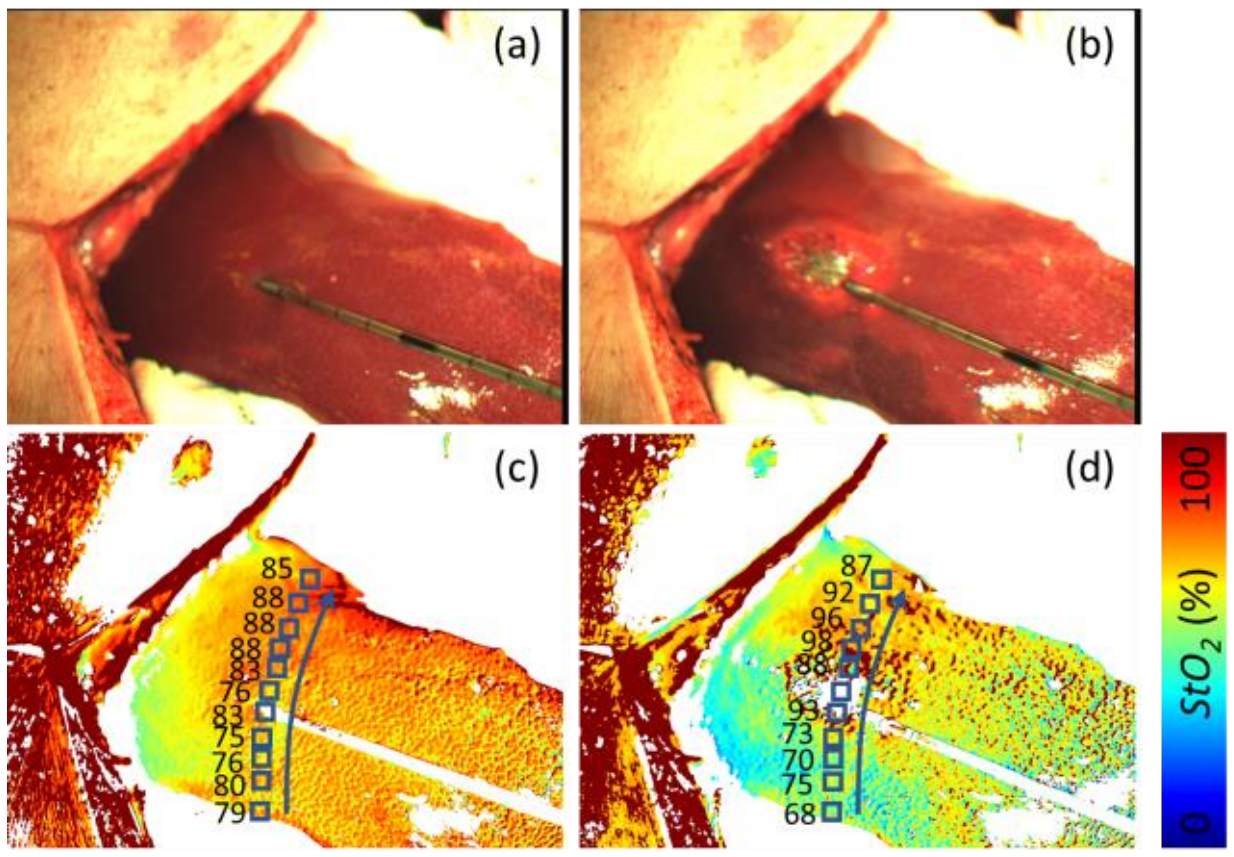

Figure 3. Spatial changes in $StO_2$ following MWA. RGB images of the liver (a) before and (b) after energy delivery. The corresponding $StO_2$ images are shown in (c) and (d). Average $StO_2$ values in a series of regions of interest are indicated in the figure.

Fig. 4 shows the interplay between temperature rise and observed changes in $StO_2$. The temperature plot shows average values within a region of interest surrounding the needle insertion point. No motion correction was performed in this experiment, therefore respiratory motion is visible in the trace as a series of spikes. Future work will investigate compensation strategies to improve this result [10]. Nevertheless it can be seen that the mean temperature in the region rises by 25°C to a peak value of 45°C over a period of two minutes. The same plot also indicates changes in mean $StO_2$ in two regions of interest, one close to the needle insertion point (ROI11) and the other in the ischaemic zone visible in Fig. 3. Baseline values vary between 80 and 95%, the variability again due in part to tissue motion during the acquisition. Despite the spatial separation of the two regions, the observed effect appears to occur at approximately the same time point, after a rise of ~10°C. At this point the ischaemic region's saturation begins to decrease, while the other region increases at a similar, but positive, rate.

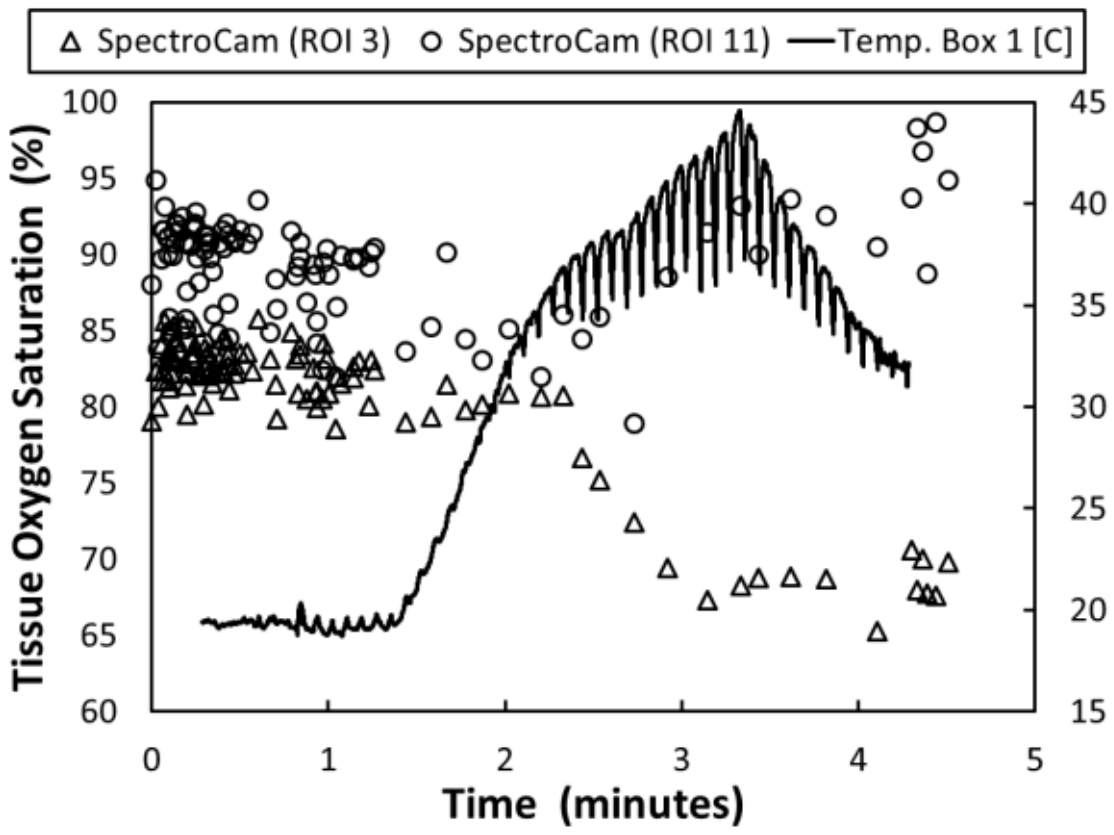

Figure 4. Variation in $StO_2$ with time in two regions of interest during application of MWA. ROI3 is ~4 cm from the point of needle insertion, while ROI11 is adjacent to it. The mean temperature from the region surrounding the needle entry point is plotted on the same axes.

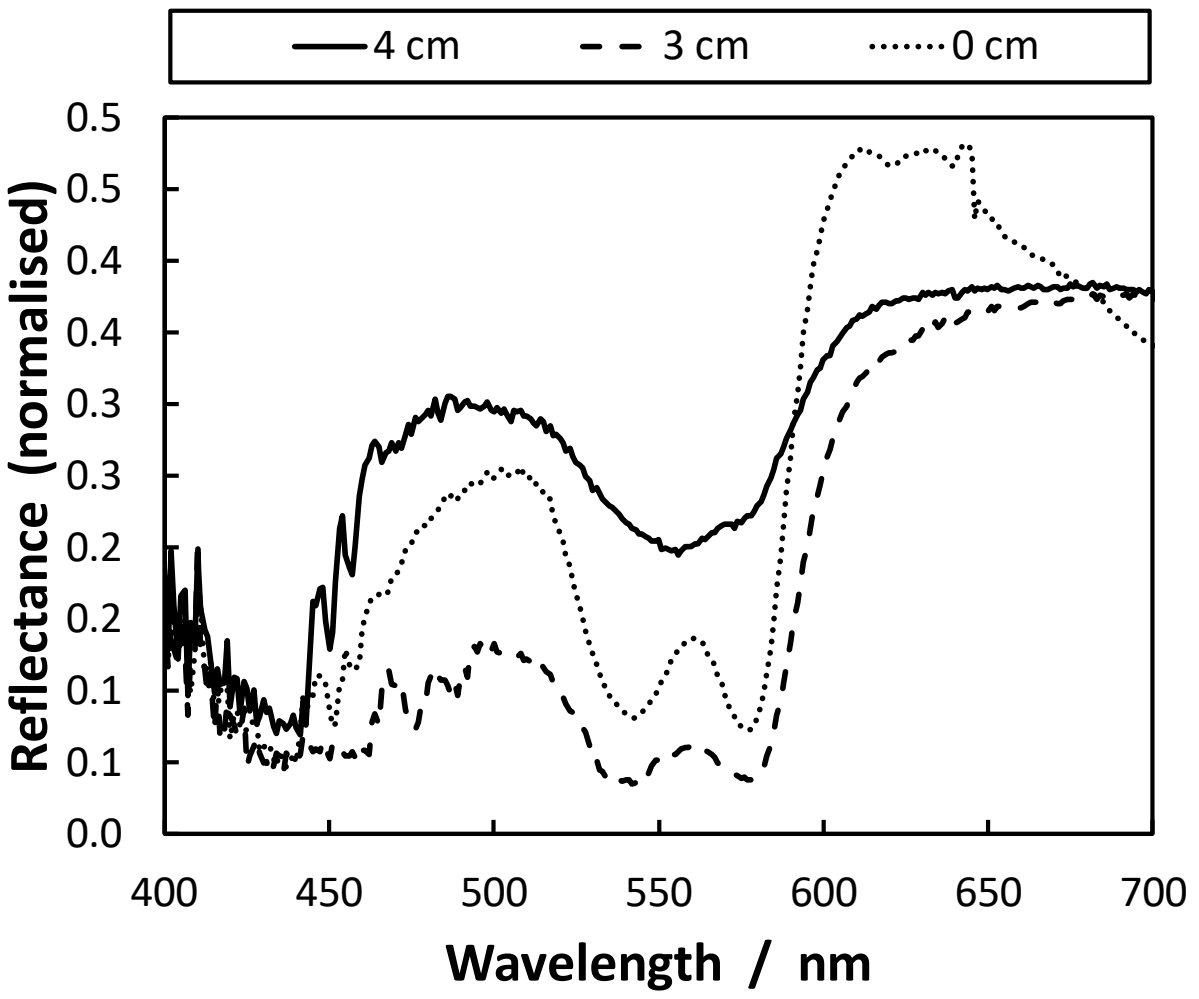

Figure 5. Diffuse reflectance spectra of liver taken at varying distances from the site of MWA needle insertion.

Fig. 5 shows diffuse reflectance spectra obtained, following MWA, from locations at varying distances from the insertion point of the needle. The region identified as ischaemic in Fig. 2 and 3 bears the distinctive single trough at approximately 550 nm associated with deoxyhaemoglobin absorption. As the probe is moved closer to the insertion point of the needle the shape of the curve begins to take on more prominent features of oxyhaemoglobin. This is apparent in the double-peak at 540/575 nm and the increasing steepness of the curve between approximately 575 and 610 nm. The shape of the high resolution reflectance spectra shows good agreement with the imaging results when compared in a similar fashion to the colour checker experiment (<10% rms), as indicated in Fig. 6.

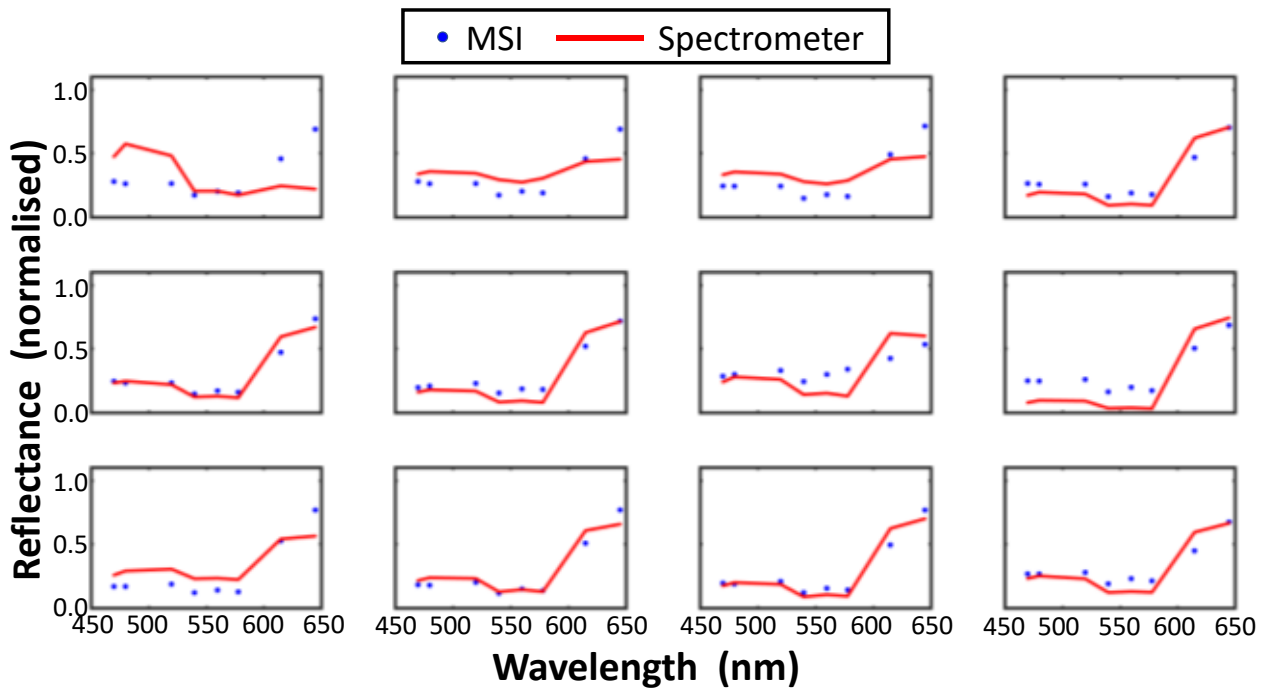

Figure 6. Plot of mean reflectance spectra recorded by SpectroCam (blue dots) against that recorded by the spectrometer (red lines). Plots correspond to the regions indicated in Fig. 3, with the top-left plot corresponding to the lower side of the organ.

## V. Discussion and Conclusions

This work presents a multimodality optical approach to studying the physiological changes that occur in tissue during application of MWA, and represents a potential method of guiding future use of the technique intraoperatively in procedures with appropriate tissue exposure.

The thermal imaging results are broadly in agreement with existing literature on the subject, which suggests that thermal destruction occurs when the local temperature exceeds 50°C.

In this experiment it was noted that the observed baseline temperature was relatively low (~20°C). Excluding the possibility of instrument calibration offset, this may be due to natural cooling due to the open incision, heat loss to the environment and vascularization resulting in the heat sink effect.

Previous literature has noted that, outside the central necrotic zone, a region of hyperaemia with increased oxygenation is expected [11]. The results presented here are also consistent with this, as evidenced by the distinctive pink 'ring' that forms concentrically about the immediate area of highest damage. An increase of up to 10% $StO_2$ was noted in this area.

Perhaps the most interesting aspect of the results displayed here is the irregular-shaped ischaemic region that spreads across one side of the liver lobe during the ablation. This has no obvious correlation with the zone of thermal damage and contradicts the trend toward higher $StO_2$ post ablation. A second interesting aspect is that this effect appears to occur synchronously with $StO_2$ increases much closer to the needle. The fact that there is no time lag between these two effects suggests that it may not be primarily due to heat damage from the needle's thermal spread, but rather a symptom of vascular damage at the central area of necrosis.

The MSI camera has been validated, in terms of its spectral accuracy, by comparing it to the results from a high resolution spectrometer. The SpectroCam was also shown to have accurately reproduced reflectance data from real liver tissue *in vivo*. The spectra where the highest errors were recorded corresponded to reflectance measurements where the probe contact pressure was seen to have varied widely. The presence of ambient light from other lamps in the operating theatre was also a possible contributing factor to spectral errors. Strong xenon light may have penetrated the tissue adjacent to the reflectance probe and been detected. The strength of this interaction could not be controlled and, following video recordings of the acquisitions, may have had a significant effect in some of the recorded spectra.

However, in spite of the presence of some outliers, the diffuse reflectance spectra provided additional information on the optical properties of the tissue. Firstly, the trend from low oxygenation in the ischaemic zone, post ablation, to increased $StO_2$ closer to the needle was supported. In the necrotic zone itself, where the MSI camera had failed, due to strong specular highlights, the contact nature of the probe allowed it to collect meaningful spectra, which also indicated high oxygenation.

Future work will employ a modified model of light-tissue interaction in order to extract more physical measures of structural changes, such as changes in scattering coefficients. Quantitative approaches [12] would allow estimation of optical scatterer size and distribution and overcome some of the limitations of the assumptions described in Section II. Our MSI camera uses visible wavelengths, which is convenient for using standard clinical light sources but limits penetration depth to approximately 2 mm. This could be extended by up to an order of magnitude by using near infrared optics.

The type of imaging approach described here may also be used to monitor other types of ablation procedures, where damage to critical structures may not be readily apparent. In this way optical measurement technologies may be incorporated as part of a feedback mechanism to monitor thermal damage to the tissue. Additional experimental work will be required to fully characterize the haemodynamic response and zone of damage at varying power settings. This will be critical as clinicians seek to explore the limits of standard and new ablation technologies in future applications while exposing the patient to the lowest possible risk.

## Acknowledgments

The authors thank Northwick Park Institute for Medical Research (NPIMR) for assistance with the surgical experiment (Home Office licence PDF66C049). The Solero microwave ablation unit was provided by AngioDynamics.